\documentclass[conference]{IEEEtran}
\IEEEoverridecommandlockouts 
    
\usepackage{cite}
\usepackage{amsmath,amssymb,amsfonts}
\usepackage{algorithmic}
\usepackage{graphicx}
\usepackage{textcomp}
\usepackage{xcolor}
\def\BibTeX{{\rm B\kern-.05em{\sc i\kern-.025em b}\kern-.08em
    T\kern-.1667em\lower.7ex\hbox{E}\kern-.125emX}}

\usepackage{comment}
\usepackage{siunitx}
\usepackage{booktabs}
\usepackage[hyphens]{url}
\usepackage{hyperref}
\usepackage{cleveref}
\usepackage{subfigure}
\usepackage{pgfplots}
\pgfplotsset{compat=newest}
\pgfplotsset{plot coordinates/math parser=false}
\newlength\figureheight
\newlength\figurewidth
\graphicspath{{./img/}}
\usetikzlibrary {decorations.fractals,spy}

\usepackage{lipsum}
\newcommand{\nota}[1]{\textcolor{blue}{#1}}

\let\olip\lipsum
\renewcommand{\lipsum}[1][]{\nota{\olip[#1]}}
\newcommand{\minus}{\scalebox{0.75}[1.0]{$-$}}

\begin{document}
\title{A Deep Learning Approach in RIS-based Indoor Localization
\thanks{This work has been supported by the Smart Networks and Services Joint Undertaking (SNS JU) under the European Union's Horizon Europe research and innovation programme under Grant Agreement No 10109710 (TERRAMETA).}}
\author{Rafael~A.~Aguiar\IEEEauthorrefmark{1}\IEEEauthorrefmark{2}, Nuno Paulino\IEEEauthorrefmark{1}\IEEEauthorrefmark{2} and Luís~M.~Pessoa\IEEEauthorrefmark{1}\IEEEauthorrefmark{2}\\
\IEEEauthorblockA{\IEEEauthorrefmark{1}INESC TEC, Porto, Portugal}
\IEEEauthorblockA{\IEEEauthorrefmark{2}Faculdade de Engenharia, Universidade do Porto, Portugal\\
\{rafael.a.aguiar, nuno.m.paulino, luis.m.pessoa\}@inesctec.pt}}


%

\maketitle

\begin{abstract}
In the domain of RIS-based indoor localization, our work introduces two distinct approaches to address real-world challenges. The first method is based on deep learning, employing a Long Short-Term Memory (LSTM) network. The second, a novel LSTM-PSO hybrid, strategically takes advantage of deep learning and optimization techniques. Our simulations encompass practical scenarios, including variations in RIS placement and the intricate dynamics of multipath effects, all in Non-Line-of-Sight conditions.
Our methods can achieve very high reliability, obtaining centimeter-level accuracy for the $98^{th}$ percentile (worst case) in a different set of conditions, including the presence of the multipath effect. Furthermore, our hybrid approach showcases remarkable resolution, achieving sub-millimeter-level accuracy in numerous scenarios.
\end{abstract}


\section{Introduction}\label{sec:intro}

Reconfigurable Intelligent Surfaces (RISs) are a transformative technology in wireless communications. They are composed of electromagnetic elements based on metasurfaces that can be individually programmed to control the wireless propagation environment \cite{tang_wireless_2021,Renzo2019SmartRE}. These elements, often called unit cells, can manipulate the phase and amplitude of reflected electromagnetic waves. This characteristic, in addition to being low-cost, high-efficiency solutions \cite{huang_reconfigurable_2019}, makes RISs a promising solution for the future of communications with 6G \cite{basharat_exploring_2022}.

In the context of localization, Smart Radio Environments enabled by RISs are promising for improving localization accuracy by addressing complex scenarios such as NLOS (Non-Line-of-Sight) and multipath effect. This is achieved by using the known position of an RIS to assist in positioning user equipment (UE). RIS-based localization systems offer significant advantages, especially in the context of \textit{Beyond-5G} networks, where the demand for accuracy extends to the millimeter level range \cite{de_lima_convergent_2021}. This is attributed to their remarkable capability to achieve very high precision.

This paper is organized as follows: \Cref{sec:sart} summarizes existing approaches on RIS based localization; \Cref{sec:model} presents our theoretical framework and model; \Cref{sec:approach} details our methodology for addressing the problem of RIS-based localization; \Cref{sec:experiments} describes the experimental setup and analyses our simulation results; finally, \Cref{sec:conclusion} concludes the paper.

\section{Related Work}\label{sec:sart}

This section presents a brief analysis of state-of-the-art RIS-based localization approaches.
Recently, several studies have directed their focus toward exploring RIS-based systems. Among these, some have specifically focused on developing algorithms for RIS-aided localization systems or improving localization accuracy for said systems. Dardari et al. \cite{Dardari_LOS_2022} proposed two algorithms for indoor localization under near-field operation and assuming NLOS between a receiver and transmitter, achieving accuracy in the millimeter range in an area of $8\times$\SI{9}{m^2}. Zhang et al. \cite{zhang_metalocalization_2021} focused on developing an algorithm that optimizes RIS phase configuration, achieving an accuracy of \SI{6}{cm} in a search space of \SI{1}{m^3}. Keykhosravi et al. \cite{keykho_siso_2022} proposed a method that estimates the position of a moving target by also estimating its radial velocity and compensating during the estimation. The method achieved accuracy in the sub-centimeter range, and they also showed that it's possible to achieve high accuracy for both static and mobile devices. Additionally, some papers have also implemented AI techniques in this context. Nguyen et al. \cite{nguyen_reconfigurable_2020} studied how the localization error in a RIS-based system behaved employing genetic algorithms (GA), neural networks (NN), and k-nearest neighbor (k-NN), showing that feature selection with k-NN yields the best results. Zhang et al. \cite{zhang_toward_2022} employed an NN to optimize the phase configuration of the RIS, achieving a localization error in the centimeter range in a space of \SI{0.125}{m^3}. This approach is focused on improving beamforming. In our previous work \cite{rafael_optimization2023}, we have also proposed one approach that leveraged Particle Swarm Optimization (PSO) and GA to estimate UE position, achieving an $85^{th}$ percentile of \SI{5}{mm} regarding localization error, where we considered an area of \SI{72}{m^2}. In this paper, we evaluate the use of NNs to improve localization for the same operating model and demonstrate improvements in terms of reliability, marking a significant step toward achieving robust and accurate UE position estimation.

\section{System Model}
\label{sec:model}
This study addresses the challenge of narrowband localization of the UE position in NLOS conditions using RIS in a downlink scenario. In this setup, the UE estimates its position by capitalizing on reflections from the RIS. The communication system adopts Orthogonal Frequency Division Multiplexing (OFDM) for signal transmission and reception. The main objective is to assess the efficacy of RIS in improving the precision of UE localization, specifically in NLOS scenarios within a downlink framework and OFDM-based communication system.

\begin{figure}[b]
    \centering
\includegraphics[width=0.5\linewidth]{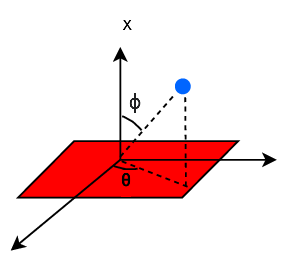}
    \caption{Tile's Coordinate System}
    \label{fig:coordsys}
\end{figure}

Our system model is based on the one proposed by Dardari et al. \cite{Dardari_LOS_2022}. In this model, we consider a large 1D RIS, composed of $K$ tiles. We consider that due to the large dimensions of the RIS, it is operating under the near-field propagation regime as a whole, however, each tile will operate under the far-field propagation regime. Each tile is separated from each other by a distance $d$, and they are composed of $N_e = N_x \times N_y$ unit cells. Here, we consider a configuration where each cell of a given tile receives the same input, resulting in the same ideal discrete phase.
The BS sends $T$ OFDM pilot symbols represented by $x_t$ with $t = 1...T$. We can select the phase response of each tile \emph{a priori} to the execution, so it's possible to define a predetermined sequence of phases for each pilot symbol. The phase response of a given cell for a specific pilot symbol is expressed as $\psi_{t,k}$.
The solid angles, illustrated in \Cref{fig:coordsys}, for each tile in relationship to the UE and BS are expressed as $\Theta_k = \{\theta_k,\phi_k\}$, where $k = 1...K$. In addition to these angles, the normalized power radiation per unit cell can be defined as \cite{Dardari_LOS_2022,dardari2021using}:
\begin{align}
    F(\Theta)= \begin{cases}\cos ^q(\theta) & \phi \in[0,2 \pi),\theta \in[0, \pi / 2] \\ 0 & \text { otherwise }\end{cases}
\end{align}

Given this, the reflection coefficient of each tile $k$ for a given pilot symbol $t$ can be expressed as: 
\begin{align}
    \beta_{t,k} = \sqrt{F(\Theta^{i}_k) F(\Theta^{r}_k)} AF(\Theta^{i}_k) AF(\Theta^{r}_k) G_c e^{j\psi_{t,k}}
\end{align}

With the superscripts $i$ and $r$ referring to the incident angle (BS-RIS) and the reflected angle (RIS-UE), respectively, and the array factor $AF(\Theta)$ modeled as \cite{Dardari_LOS_2022,balanis2016antenna}:
\begin{align}  
    AF(\Theta) = \frac{1}{\sqrt{N_e}}\frac{\sin(\frac{\pi N_x d_x}{\lambda}\sin\theta \sin \phi)}{\sin(\frac{\pi d_x}{\lambda}\sin\theta \sin \phi)} \frac{\sin(\frac{\pi N_y d_y}{\lambda}\sin\theta \cos \phi)}{\sin(\frac{\pi d_y}{\lambda}\sin\theta \cos \phi)}
\end{align}

In this equation, $d_x$ and $d_y$ represent the distances in each axis between cells, and for a boresight cell gain ($G_c$) of \SI{5}{dBi}, the associated normalized power radiation parameter ($q$) is set to 0.57 \cite{Dardari_LOS_2022}.

Finally, we can represent the signal measured by the UE (in NLOS) for the pilot symbol $t$ as:

\begin{align}
\label{eq:meas}
y_t=x_t f^{(\mathrm{mp})} + x_t\sum_{k=1}^K \beta_{t, k} h_k^{(\mathrm{r})}+w_t
\end{align}

Where $w_t\sim \mathcal{CN}(0, \sigma^2)$ models the measurement noise, with $\sigma^2$ as the noise power. Furthermore, we can also define $x_t = 1, \forall t = 1...T$, since the symbol is known at the UE. In this equation, $h_k^{(\mathrm{r})}$ is the cascaded channel coefficient (BS-RIS-UE), and $f^{(\mathrm{mp})}$ is the multipath component of the direct link, which will be zero in the case where this effect is not considered. Both components are explained thoroughly in \cite{Dardari_LOS_2022}.

\section{Proposed Approach}\label{sec:approach}

In this section, we delve into the details of our methodology, highlighting the sequential steps involved in utilizing deep learning to improve localization accuracy further.

The \emph{Direct Position} algorithm employs the following cost function minimization:

\begin{align}
\label{eq:cost}
    \hat{\mathbf{p}}=\arg \min _{\left\{\mathbf{p}, \phi_0\right\}} \sum_{t=1}^T \frac{\tilde{a}_{t}^2}{\sigma^2} \sin ^2\left(\tilde{\phi}_{t}-\phi_t(\bold p)- \phi_0\right)
\end{align}
With $\tilde{a}_{t}$ and $\tilde{\phi}_{t}$ being the magnitude and phase of the measured signal $y_t$, $\phi_0$ is the phase offset between the BS and UE, which is also a parameter to be estimated. 
Finally,  $\phi_t(\bold p)$ is the phase of pilot symbol t, which depends only on the true UE position $p$. Therefore, estimating the UE position is equivalent to minimizing \Cref{eq:cost} as a function of the set of phases $\tilde{\phi}_{t}$, since this set has a univocal relationship to the true position.
This hypothetical phase can be calculated as:
\begin{align}
\phi_t(\mathbf{p})=\arg \left(\sum_{k=1}^K \beta_{t, k} \cdot h_{n, k}^{(\mathrm{dp})}(\bold p,\phi_0)\right)
\end{align}

In this equation, $h_{n, k}^{(\mathrm{dp})}(\bold p,\phi_0)$ is the cascaded channel of \Cref{eq:meas}, when not considering the multipath effect. Dardari et al. \cite{Dardari_LOS_2022} factored out $\bold p,\phi_0$ to show that the cascaded channel is a function of these unknown variables.

The goal is to use the signal measured by the UE and the reflection coefficients of the RIS tiles to estimate the UE position. Since we want to obtain a position in two axes where each axis has a predetermined range, this can be modeled as a regression problem. The choice of regression technique now depends on the data being processed.

Neural networks are composed of neurons connected by weights. Each neuron produces an output by performing a linear combination of their inputs and weights and adding a bias. To introduce non-linearity, the output is then passed through an activation function. The simplest models of NN are called Feed Forward Neural Networks (FFNN). This architecture is often insufficient to predict complex patterns, especially in regression problems \cite{looney1997pattern,bataineh_neural_2017}. Another widely used architecture type is the Convolution Neural Network (CNN). CNNs are composed of convolutional layers, pooling layers, and fully connected layers. Even though these networks are very powerful, they are usually applied in the context of computer vision since they are better suited to image data \cite{liu_survey_2021,simonyan_very_2015}. Another type of network is called Recurrent Neural Network (RNN). These networks are used when the data is affected by a sequential dependency. Applications include financial data analysis and speech pattern recognition. These networks use cyclic connections, constructing a type of memory \cite{eck_neural_2018,azzouni_long_2017}. The main limitation of these models is that they suffer from vanishing gradients. The vanishing gradient is a phenomenon that occurs during the training of deep neural networks using the backpropagation algorithm. When the gradients become too small, updating the network's parameters may become impossible or inefficient. To address this issue, Hochreiter et al. \cite{hochreiter_long_1997} proposed the Long Short-Term Memory (LSTM) model. Considering that the input intended for this model follows a sequence, we considered that LSTM was the fittest model.

\subsection{Data Processing}
The dataset was generated in a simulation environment where the UE positions (the target outputs for the model) were randomly drawn from a uniform distribution. For each position, the reflection coefficient and the corresponding measured signal were recorded. This process was repeated $10^5$ times, resulting in a highly diverse dataset, making data augmentation techniques unnecessary.
For every sample, the inputs $y_{t}$, $\beta_{t,k}$, can be represented as complex matrices of sizes $T\times1$ and $T \times K$, respectively.
Although NN implementations dealing with complex numbers exist \cite{hirose_nature_2011,wolter_complex_2018}, this is not the standard case. This is due to the added complexity of the training and generally because using real numbers has been very well established.
To address this issue, each matrix was divided into two matrices corresponding to the phase and magnitude of the original values.
Subsequently, the four inputs comprising phase and magnitude information were normalized to conform to a $\mathcal{N}(0, 1)$ distribution. In addition to normalizing the inputs, we also decided to normalize the outputs ($x,y$). This decision ensures a consistent scale for both output dimensions, promoting equal weighting of importance by the neural network.
Following this step, all the input tensors were concatenated. Finally, the dataset was split into 80/10/10 (training, validation, and testing), and to improve the NN's generalization capability, we divided the dataset into batches of 32 samples.

\subsection{Model}

The proposed architecture begins with a flattening layer to transform the input into a one-dimensional tensor. Subsequently, a bidirectional LSTM layer is employed with an input size of $2 \cdot T\times (K+1)$, a hidden size of 500. This bidirectional LSTM layer enables the model to capture both forward and backward temporal dependencies, enhancing its ability to discern intricate patterns within the sequential data. To reduce overfitting, we added a dropout rate of $40\%$ in this layer. This layer produces an output of 1000 $(hidden size \times 2)$ and is responsible for detecting features, followed by a sequence of fully connected layers, contributing to the feature extraction process. The subsequent fully connected layers have dimensions $(1000\times2048)$, $(2048 \times 512)$, $(512 \times 64)$, respectively. Nonlinearity is added using the ReLU activation function $f(x) = \max(0, x)$
Finally, the model concludes with a final fully connected layer with two output neurons using a linear activation function suitable for regression tasks. 
The architecture is implemented and executed on a \emph{NVIDIA GeForce GTX 970} using the cuda package in PyTorch. Since the model learns to predict a position by analyzing tens of thousands of positions in the room, this method can be considered as a fingerprinting-based technique.

\begin{table}[tb]
  \centering
  \caption{Model Summary}
  \label{tab:model_summary}
  \begin{tabular}{lcc}
    \toprule
    \textbf{Layer} & \textbf{Output} & \textbf{Activation} \\
    \midrule
    Input (Flatten) & $2 \cdot T\times (K+1)$ & - \\
    LSTM (Bidirectional) & 1000 & - \\
    FC1 & 2048 & ReLU \\
    FC2 & 512 & ReLU \\
    FC3 & 64 & ReLU \\
    Output (FC4) & 2 & Linear \\
    \bottomrule
  \end{tabular}
\end{table}

\subsection{Training}
In terms of training, the Mean Squared Error (MSE) loss is utilized, and the model parameters are updated using the Adam optimizer \cite{kingma_adam_2017} with a learning rate of $0.001$.
The training process involves iterating through a specified number of epochs, and the model is updated after iterating through all the samples in a batch. After each epoch, the model's performance is evaluated on the validation dataset. The learning rate is dynamically adjusted using a scheduler tied to the validation loss. It undergoes a $20\%$ reduction when the validation loss does not improve for 10 consecutive epochs. The training loop involves setting the model to training mode, calculating and backpropagating the loss, optimizing the model parameters, and then switching to evaluation mode for validation. The model with the best validation loss is saved. The model is saved based on the best validation loss rather than the training loss, as the two events may not coincide. This practice ensures the retention of the model that exhibits the best performance on unseen data.

\subsection{LSTM-PSO Hybrid}
\begin{figure}
    \centering
\includegraphics[width=0.7\linewidth]{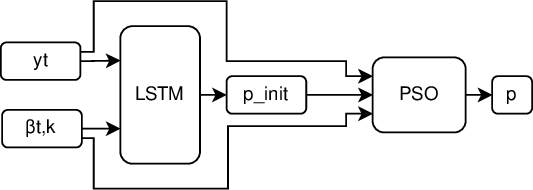}
    \caption{\emph{LSTM-PSO Hybrid}}
    \label{fig:lstmpso_hybrid}
\end{figure}

After training and testing the deep learning model, we concluded that it has a good capability of producing estimations reliably near the true position ($98^{th}$ percentile below \SI{22}{cm}), which is a significant improvement over our previous work \cite{rafael_optimization2023}. On the other hand, the model couldn't achieve very high resolution.
Considering this, we developed a new algorithm. The algorithm operates by initially providing the neural network with two inputs. Subsequently, the outputs undergo a reverse normalization transform to ensure they align with the scale of the possible positions of the UE. Using this initial estimate, the algorithm defines a neighborhood around the computed point, with the neighborhood size being a tunable parameter. The final step involves minimizing the cost function (\Cref{eq:cost}) using PSO, where the optimization boundaries are set to the neighborhood of the initial estimate.
This method enables the algorithm to yield high-resolution solutions due to the bounded global optimization while simultaneously instilling a high degree of confidence in the obtained solution's accuracy and reliability.
\Cref{fig:lstmpso_hybrid} illustrates the architecture of the proposed algorithm.

\section{Experimental Evaluation}\label{sec:experiments}

\subsection{Experimental Setup}

\begin{figure}[!b]
    \centering
\includegraphics[width=0.7\linewidth]{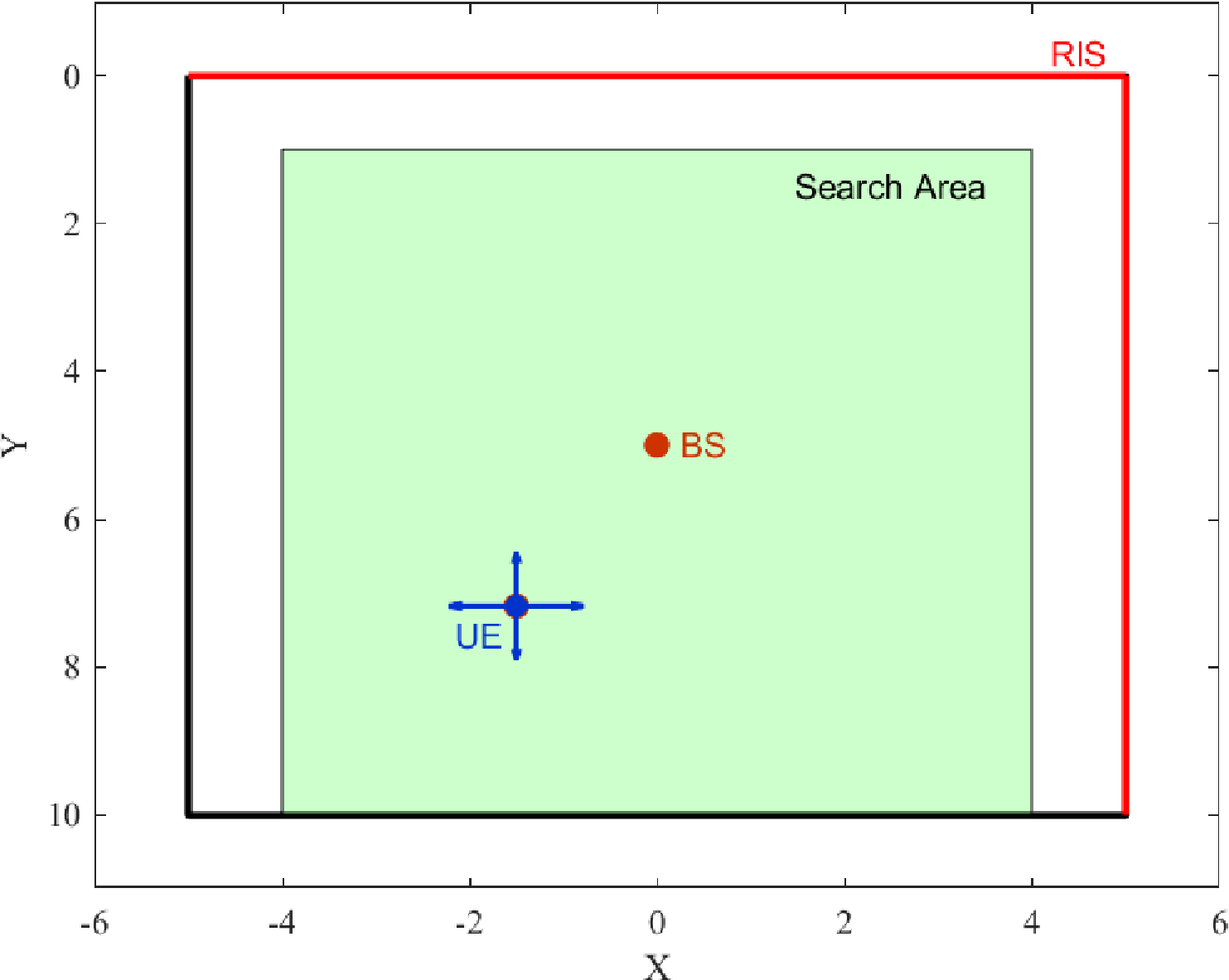}
    \caption{Top-View Room Configuration \cite{rafael_optimization2023}}
    \label{fig:room}
\end{figure}

The experimental setup is the same as used in the previous work \cite{rafael_optimization2023} to ensure a fair comparison between the algorithms.
We consider a room with dimensions $10 \times 10 \times 3$ \SI{}{m^3}. From now on, the height of the BS and RIS will be represented by the same symbol, $z$, because they will always be in the same plane. The BS will be positioned in the center of the room with $p_{TX} = (0,5,z)$, with $z$ variable.
The large RIS will be divided into two segments. The position of the tiles of the first segment is expressed as $p_k = (\minus5, 0, z)$ to $(5, 0, z)$ and the tiles from the second segment $p_k = (5, 0, z)$ to $(5, 10, z)$, again with $z$ variable and a \SI{20}{cm} distance between consecutive tiles' centroids. The system operates at \SI{3.5}{GHz} with $T = 32$ OFDM pilot symbols. We also implement narrowband localization, with a single pilot subcarrier and 2048 subcarriers, and consider a noise power $\sigma^2 =$ \SI{-120.2}{dBm}. Regarding the RIS architecture, we implemented $K = 100$ tiles, each comprising $4 \times 25$ unit cells. The UE position $\textbf{p}$ will be drawn according to a uniform distribution from the following region $(\minus 4,1)$ to $(4,10)$ in the plane $z = 1$. The physical configuration is illustrated in \Cref{fig:room}.

\subsection{Results}
In this section, we will provide the results for the performance of the algorithms (LSTM and \emph{LSTM-PSO Hybrid}) for different conditions, as well as compare them against the previously developed methods. Initially, we tested the NLOS scenario between BS and UE, and the most simple case was where the multipath effect was not considered and the RIS and BS were on the same plane as the UE ($z=1$).

\Cref{tab:model_comparison_regression} compares NN architectures using the testing dataset. This comparison aimed to prove that the LSTM, particularly a Bidirectional LSTM, is the most suitable architecture to implement as the ML block in our proposed algorithm. An important consideration is that the latency of the models is in the millisecond range, which is a significant improvement from any brute-force or even our previously developed optimization algorithms (more than $10000$ times faster).
\begin{table}[tb]
  \centering
  \caption{NN Comparison}
  \label{tab:model_comparison_regression}
  \begin{tabular}{lcccc}
    \toprule
    \textbf{Model} & \textbf{Loss (MSE)} & \textbf{Params} & \textbf{Epochs} & \textbf{Latency} \\
    \midrule
    LSTM Bidirectional & 0.0003 & \SI{31.0}{M} & 25 & \SI{5.4}{ms} \\
    LSTM & 0.0006 & \SI{16.0}{M} & 25 & \SI{5.3}{ms} \\
    RNN & 0.001 & \SI{5.6}{M} & 25 & \SI{4.4}{ms} \\
    MLP & 0.0007 & \SI{3.4}{M} & 25 & \SI{4.5}{ms} \\
    \bottomrule
  \end{tabular}
\end{table}

\begin{figure}[!tb]
    \centering
    \subfigure[LSTM]{
        \includegraphics[width=0.45\linewidth]{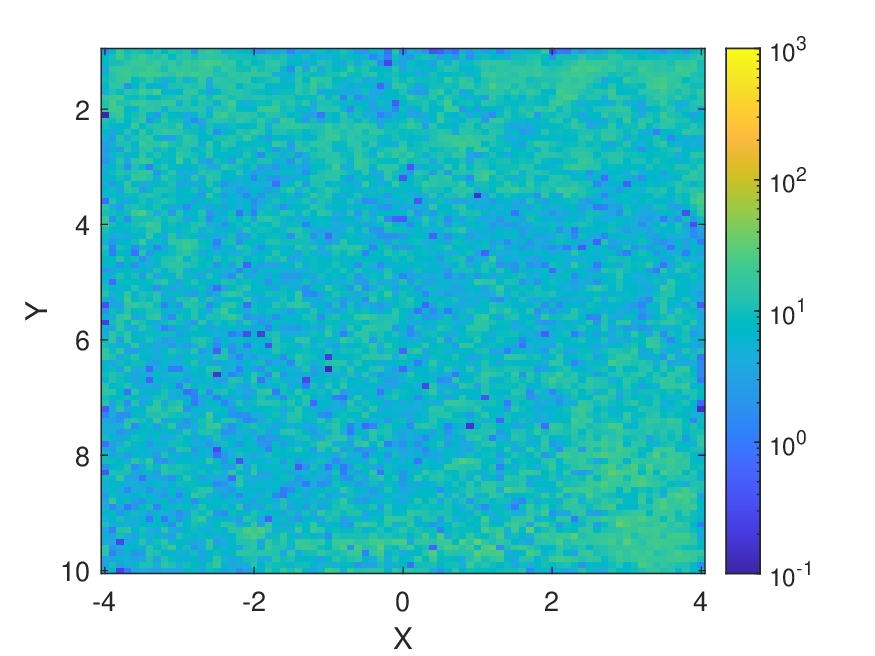}
        \label{subfig2:a}
    }
    \hfill
    \subfigure[\emph{LSTM-PSO Hybrid}]{
        \includegraphics[width=0.45\linewidth]{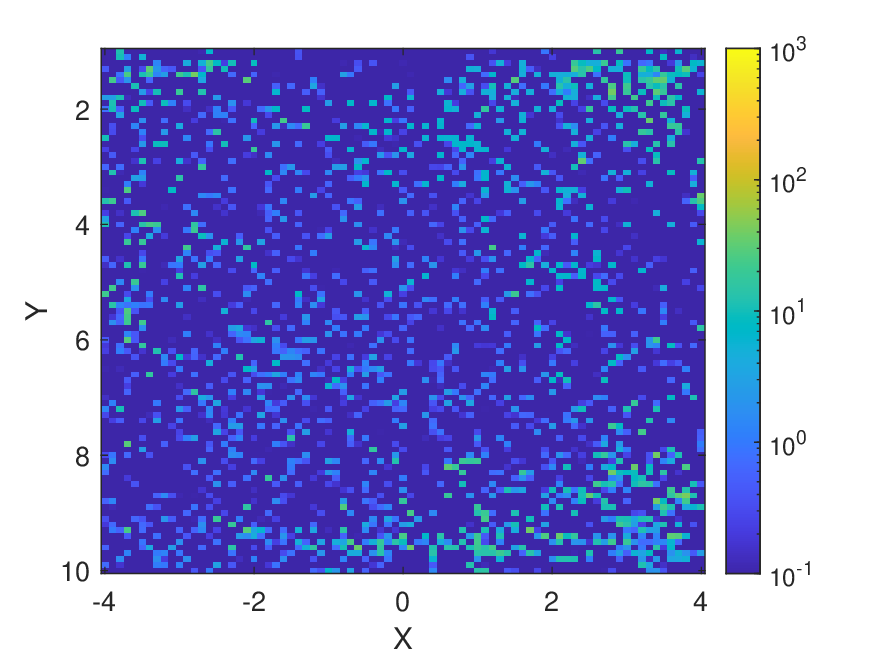}
        \label{subfig2:b}
    }
    \hfill
    \caption{Localization error [cm] heatmaps at 1-meter height}
    \label{fig:heatmapsz1}
\end{figure}

To gain a comprehensive understanding of the model's performance, we generated a dataset that captures all possible points of the search space with a resolution of 10 cm, resulting in 7371 simulations. The heatmap in \Cref{fig:heatmapsz1} illustrates the distribution of errors across the dataset, providing a holistic overview of the model's behavior with respect to the true positions. We observe that the LSTM heatmap does not show any peaks in error, which means that it is very good at obtaining at least sub-optimal solutions. On the other hand, the proposed \emph{LSTM-PSO Hybrid} also does not show large peaks in error and shows very low error values (below \SI{1}{mm}) in most of the area. We can also check that the algorithm tends not to converge on the optimal solution (errors between \SI{1}{}-\SI{10}{cm}) near the top right and bottom corners, which is something already observed in \cite{rafael_optimization2023}, and attributed to the narrow field-of-view of the UE with respect to the RIS.

\begin{figure}[tb]
  \centering
  \resizebox{0.8\columnwidth}{!}{%
%
%
\definecolor{mycolor1}{rgb}{0.92941,0.69412,0.12549}%
\definecolor{mycolor2}{rgb}{0.30196,0.74510,0.93333}%
\definecolor{mycolor3}{rgb}{0.63529,0.07843,0.18431}%
\definecolor{mycolor4}{rgb}{0.68627,0.74510,0.99608}%

\begin{tikzpicture}
[spy using outlines={rectangle, magnification=2.5, height = 4, width = 2, connect spies}]

\begin{axis}[%
width=4.521in,
height=3.566in,
at={(0.758in,0.481in)},
scale only axis,
xmin=10,
xmax=100,
xlabel style={font=\color{white!15!black}},
xlabel={Percentile},
ymode=log,
ymin=5.59605500819389e-06,
ymax=55960.5500819386,
yminorticks=true,
ylabel style={font=\color{white!15!black}},
ylabel={Localization Error [cm]},
axis background/.style={fill=white},
xmajorgrids,
ymajorgrids,
yminorgrids,
legend style={at={(0.03,0.97)}, anchor=north west, legend cell align=left, align=left, draw=white!15!black}
]
\addplot [color=mycolor1, line width=1.5pt, mark size=2pt, mark=x, mark options={solid, mycolor1}]
  table[row sep=crcr]{%
10	5.68432933633599e-05\\
25	0.000131694599775132\\
40	0.000311237326706081\\
50	0.00252900897175946\\
60	104.433255560446\\
65	247.011122428152\\
70	339.469190754969\\
75	366.680668516901\\
80	392.425345197845\\
85	424.204088278775\\
90	529.266478302717\\
92.5	616.622127336709\\
95	686.075020251337\\
98	795.786647377898\\
};
\addlegendentry{PSO}

\addplot [color=green, line width=1.5pt, mark size=2pt, mark=square, mark options={solid, green}]
  table[row sep=crcr]{%
10	0.707106781186595\\
25	1.52970585407789\\
40	2.5961509971495\\
50	3.62353418639874\\
60	4.96487663492262\\
65	5.71401785086466\\
70	6.52809307190281\\
75	7.52213333908957\\
80	9.20727709365784\\
85	13.0013076149386\\
90	43.4305301261686\\
92.5	158.055332785984\\
95	380.840027935676\\
98	472.312769581936\\
};
\addlegendentry{Discrete GA}

\addplot [color=red, line width=1.5pt, mark size=2pt, mark=o, mark options={solid, red}]
  table[row sep=crcr]{%
10	0.000109692463530687\\
25	0.000215637080555928\\
40	0.000353834579504479\\
50	0.000493301264087652\\
60	0.000707419901117674\\
65	0.000873527810738596\\
70	0.00111751265331327\\
75	0.00156119040649538\\
80	0.00421402519640526\\
85	276.459612119433\\
90	381.210037692997\\
92.5	406.492940875341\\
95	480.579415121642\\
98	706.49117139191\\
};
\addlegendentry{A Priori PSO}

\addplot [color=mycolor2, line width=1.5pt, mark size=2pt, mark=asterisk, mark options={solid, mycolor2}]
  table[row sep=crcr]{%
10	0.000116267972465152\\
25	0.000228782108327731\\
40	0.000367687071325794\\
50	0.000509459979271721\\
60	0.000736627292183654\\
65	0.000893412373950041\\
70	0.00113435441081264\\
75	0.00159245430130146\\
80	0.00336184139703538\\
85	0.449237553505845\\
90	29.0722273553288\\
92.5	157.911463700067\\
95	381.102913102156\\
98	438.561944702087\\
};
\addlegendentry{GA-PSO Hybrid}

\addplot [color=black, line width=1.5pt, mark size=2pt, mark=triangle, mark options={solid, rotate=90, black}]
  table[row sep=crcr]{%
10	0.000182187714598709\\
25	0.000377587486676053\\
40	0.000643403065322612\\
50	0.000946629562205778\\
60	0.00152808392740616\\
65	0.00217337431612471\\
70	0.00414456669479894\\
75	0.10167759757372\\
80	0.316604725108972\\
85	0.735013621924\\
90	1.8808081959388\\
92.5	4.01866470644153\\
95	6.69055880030207\\
98	14.0843324466367\\
};
\addlegendentry{LSTM-PSO Hybrid}

\addplot [color=mycolor3, line width=1.5pt, mark size=2pt, mark=triangle, mark options={solid, rotate=270, mycolor3}]
  table[row sep=crcr]{%
10	3.03639384639808\\
25	5.10132575187349\\
40	7.01387439849882\\
50	8.20983886718736\\
60	9.45971108904057\\
65	10.2222368682857\\
70	10.9843336095202\\
75	11.8784240329698\\
80	12.8925500422723\\
85	14.1458789114603\\
90	15.7758535207719\\
92.5	17.0027985841769\\
95	18.492544912403\\
98	21.3557436144968\\
};
\addlegendentry{LSTM}

\addplot [color=mycolor4, line width=1.5pt, mark size=2pt, mark=square, mark options={solid, mycolor4}]
  table[row sep=crcr]{%
10	27.4\\
25	38\\
40	72.8\\
50	127\\
60	179.4\\
65	240.2\\
70	280.3\\
75	306\\
80	341.9\\
85	392.4\\
90	410.1\\
92.5	445.2\\
95	520.9\\
98	688.9\\
};
\addlegendentry{Gradient Descent}
\end{axis}

\spy [black, height = 6cm,width = 3cm] on (12.7,7.6)
              in node [right] at (11.8,2.5);
              
\end{tikzpicture}%
  }
  \caption{Localization Error Distribution along Percentiles}
  \label{fig:fullcomp}
\end{figure}
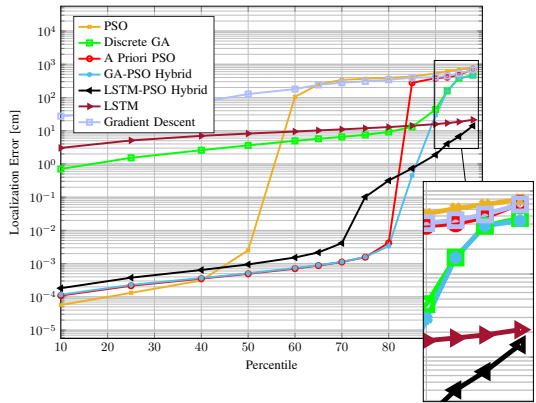

\Cref{fig:fullcomp} shows the performance of the two methods developed in this paper against our previously developed solutions. We can observe that the LSTM model is very reliable, with a $98^{th} $ percentile of \SI{22}{cm}. This is significantly better than the previously best method (\emph{GA-PSO Hybrid}) that obtained a $90^{th} $ percentile of \SI{29}{cm}. The main limitation of the model is not being able to achieve very high resolution, with practically all the errors being above \SI{1}{cm}. However, while the previous method had an average estimation time of \SI{60}{s}, the proposed LSTM model has a latency around \SI{5}{ms}. Regarding the \emph{LSTM-PSO Hybrid}, we can observe even better performance. It combines the reliability of the NN and the high resolution of the optimization techniques. We observe high accuracy having an $80^{th}$ percentile of \SI{3}{mm}. However, this algorithm shows the best performance at the higher percentiles. While the previous methods suffer from completely offset estimates (due to the cost function being a probabilistic approximation), this algorithm is much more robust against outliers since it relies on the NN to capture complex patterns of the measured signals.  Finally, we also confirm that the algorithm performance is bounded by the NN performance, with a $98^{th} $ percentile error of \SI{14}{cm}, a great improvement from \emph{GA-PSO Hybrid} that shows an error of \SI{4.6}{m} for the same metric. Regarding an ablation study, we analyzed the contribution of each block separately to the \emph{LSTM-PSO Hybrid}. We confirmed that the LSTM primarily ensures high reliability, whereas the PSO block is crucial for achieving high precision. Additionally, we observed that the LSTM block alone performs well but with less precision than the Hybrid method while relying solely on the PSO algorithm independently is not viable."

\begin{figure}[!tb]
  \centering
  \resizebox{0.8\columnwidth}{!}{%
%
%
\definecolor{mycolor1}{rgb}{0.63529,0.07843,0.18431}%
\definecolor{mycolor2}{rgb}{1.00000,0.41176,0.16078}%
\definecolor{mycolor3}{rgb}{0.71765,0.27451,1.00000}%
\begin{tikzpicture}
[spy using outlines={rectangle, magnification=2, size = 3cm, connect spies}]

\begin{axis}[%
width=4.521in,
height=3.566in,
at={(0.758in,0.481in)},
scale only axis,
xmin=0,
xmax=100,
xlabel style={font=\color{white!15!black}},
xlabel={Percentile},
ymode=log,
ymin=0.0001,
ymax=100,
yminorticks=true,
ylabel style={font=\color{white!15!black}},
ylabel={Localization Error [cm]},
axis background/.style={fill=white},
xmajorgrids,
ymajorgrids,
yminorgrids,
legend style={at={(0.03,0.97)}, anchor=north west, legend cell align=left, align=left, draw=white!15!black}
]
\addplot [color=mycolor1, line width=1.5pt, mark size=2pt, mark=o, mark options={solid, mycolor1}]
  table[row sep=crcr]{%
10	0.00410156039845881\\
25	0.00829026623730807\\
40	0.0134339936874255\\
50	0.0184962609283621\\
60	0.0271611676178953\\
65	0.0367622136658335\\
70	0.0841376786636818\\
75	0.311465746783759\\
80	0.746295356716128\\
85	2.08181018449326\\
90	10.8356722688654\\
92.5	17.5560797139064\\
95	24.5681839314615\\
98	34.7435956762128\\
};
\addlegendentry{LSTM-PSO, 3 m}

\addplot [color=mycolor2, line width=1.5pt, mark size=2pt, mark=x, mark options={solid, mycolor2}]
  table[row sep=crcr]{%
10	5.05525261565188\\
25	8.39262026065241\\
40	11.2717531032063\\
50	13.2104674527517\\
60	15.3840650085042\\
65	16.6080453862381\\
70	17.9702739487276\\
75	19.4588231003722\\
80	21.3444016343437\\
85	23.4604058859048\\
90	26.6709072189254\\
92.5	29.3413071783869\\
95	32.485218492886\\
98	42.4485607223307\\
};
\addlegendentry{LSTM, 3 m}

\addplot [color=black, line width=1.5pt, mark size=2pt, mark=triangle, mark options={solid, rotate=90, black}]
  table[row sep=crcr]{%
10	0.000182187714598709\\
25	0.000377587486676053\\
40	0.000643403065322612\\
50	0.000946629562205778\\
60	0.00152808392740616\\
65	0.00217337431612471\\
70	0.00414456669479894\\
75	0.10167759757372\\
80	0.316604725108972\\
85	0.735013621924\\
90	1.8808081959388\\
92.5	4.01866470644153\\
95	6.69055880030207\\
98	14.0843324466367\\
};
\addlegendentry{LSTM-PSO, 1 m}

\addplot [color=mycolor3, line width=1.5pt, mark size=2pt, mark=diamond, mark options={solid, mycolor3}]
  table[row sep=crcr]{%
10	3.03639384639808\\
25	5.10132575187349\\
40	7.01387439849882\\
50	8.20983886718736\\
60	9.45971108904057\\
65	10.2222368682857\\
70	10.9843336095202\\
75	11.8784240329698\\
80	12.8925500422723\\
85	14.1458789114603\\
90	15.7758535207719\\
92.5	17.0027985841769\\
95	18.492544912403\\
98	21.3557436144968\\
};
\addlegendentry{LSTM, 1 m}

\end{axis}

\spy [black, height = 4cm,width = 2cm] on (12.8,9.1)
              in node [right] at (12.5,4);
\end{tikzpicture}%
  }
  \caption{Localization Error Comparison for RIS and BS height: 3m vs 1m}
  \label{fig:heightcomp}
\end{figure}
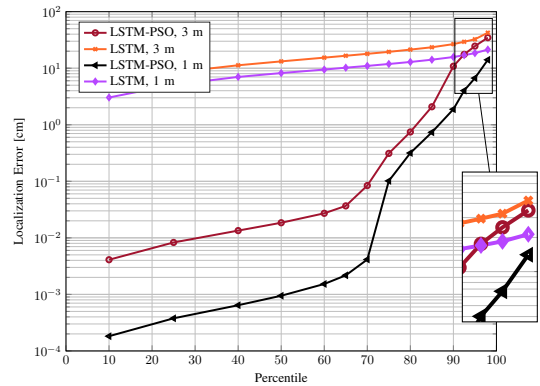

We also compared the behavior of the two methods in a more realistic scenario where the BS and the RIS are placed at \SI{3}{m} height, i.e., in the region connecting the ceiling and walls. Since the signals will differ in this configuration, the NN model was trained again with a dataset obtained for this scenario.
Analyzing \Cref{fig:heightcomp}, we observe that for the NN model, the localization error keeps in the centimeter range, with a $95^{th}$ percentile of \SI{32}{cm} \emph{vs} \SI{18}{cm} at 3 and 1-meter height, respectively.
The hybrid method shows a more pronounced difference, namely in the lower percentiles. However, at the 3-meter height, the method achieves a $70^{th}$ percentile below \SI{1}{mm}, making this difference irrelevant in practical systems. The difference is smaller for the most relevant metrics in the higher percentiles. The algorithm performs best at $z = 1$~meter with an $85^{th}$ percentile = \SI{7}{mm} \emph{vs} \SI{2}{cm} and a $95^{th}$ percentile = \SI{7}{cm} \emph{vs} \SI{24}{cm}. Even though the performance is worst for ceiling placement, the results suggest that the algorithms are still viable in these conditions.

  We also trained the same model now with a dataset obtained under the multipath regime. Curiously, the NN model does not show significant differences between both scenarios due to the ability of the NN to capture patterns even when there is multipath interference. However, this is not the case for the hybrid method. We observed a notable decrease in performance, and interestingly, the hybrid method performed worse than LSTM in both conditions. This shows that even though the LSTM can accurately determine the region where the UE is located, PSO cannot, with most estimations leading to larger errors. This can be explained because PSO works on top of the \emph{Direct Positioning} formulation, which does not take into account the multipath effect in its probabilistic estimations.

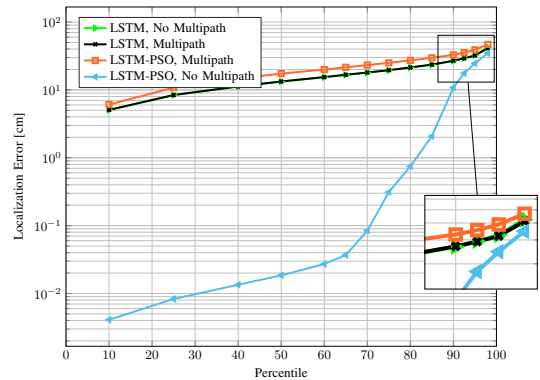
\begin{figure}[!tb]
  \centering
  \resizebox{0.8\columnwidth}{!}{%
%
%
\definecolor{mycolor1}{rgb}{1.00000,0.41176,0.16078}%
\definecolor{mycolor2}{rgb}{0.30196,0.74510,0.93333}%

\begin{tikzpicture}
[spy using outlines={rectangle, magnification=2, size = 3cm, connect spies}]

\begin{axis}[%
width=4.521in,
height=3.566in,
at={(0.758in,0.481in)},
scale only axis,
xmin=-2.08166817117217e-15,
xmax=100,
xlabel style={font=\color{white!15!black}},
xlabel={Percentile},
ymode=log,
ymin=0.00166735263919309,
ymax=166.735263919309,
yminorticks=true,
ylabel style={font=\color{white!15!black}},
ylabel={Localization Error [cm]},
axis background/.style={fill=white},
xmajorgrids,
ymajorgrids,
yminorgrids,
legend style={at={(0.03,0.97)}, anchor=north west, legend cell align=left, align=left, draw=white!15!black}
]
\addplot [color=green, line width=1.5pt, mark size=2pt, mark=triangle, mark options={solid, rotate=270, green}]
  table[row sep=crcr]{%
10	5.05525261565188\\
25	8.39262026065241\\
40	11.2717531032063\\
50	13.2104674527517\\
60	15.3840650085042\\
65	16.6080453862381\\
70	17.9702739487276\\
75	19.4588231003722\\
80	21.3444016343437\\
85	23.4604058859048\\
90	26.6709072189254\\
92.5	29.3413071783869\\
95	32.485218492886\\
98	42.4485607223307\\
};
\addlegendentry{LSTM, No Multipath}

\addplot [color=black, line width=1.5pt, mark size=2pt, mark=x, mark options={solid, black}]
  table[row sep=crcr]{%
10	5.07513628771274\\
25	8.40536259626881\\
40	11.2452397390384\\
50	13.324617640734\\
60	15.4438416562124\\
65	16.7265026212087\\
70	18.04442432251\\
75	19.5684387038369\\
80	21.4431776001548\\
85	23.6137167131319\\
90	26.9203960473916\\
92.5	29.1498734370372\\
95	32.0465663422147\\
98	41.2333747286382\\
};
\addlegendentry{LSTM, Multipath}

\addplot [color=mycolor1, line width=1.5pt, mark size=2pt, mark=square, mark options={solid, mycolor1}]
  table[row sep=crcr]{%
10	6.07725748938629\\
25	10.779358834852\\
40	14.6525535464674\\
50	17.3854292907252\\
60	19.9750028246856\\
65	21.5381614780989\\
70	23.2365150795846\\
75	25.062446578513\\
80	27.2393938331052\\
85	29.725000312379\\
90	32.8860661743469\\
92.5	35.4042870943383\\
95	38.9939231525298\\
98	46.8149437875248\\
};
\addlegendentry{LSTM-PSO, Multipath}

\addplot [color=mycolor2, line width=1.5pt, mark size=2pt, mark=triangle, mark options={solid, rotate=90, mycolor2}]
  table[row sep=crcr]{%
10	0.00410156039845881\\
25	0.00829026623730807\\
40	0.0134339936874255\\
50	0.0184962609283621\\
60	0.0271611676178953\\
65	0.0367622136658335\\
70	0.0841376786636818\\
75	0.311465746783759\\
80	0.746295356716128\\
85	2.08181018449326\\
90	10.8356722688654\\
92.5	17.5560797139064\\
95	24.5681839314615\\
98	34.7435956762128\\
};
\addlegendentry{LSTM-PSO, No Multipath}

\end{axis}


\spy [black, height = 2.5cm,width = 3cm] on (12.6,8.9)
              in node [right] at (11.5,4);
\end{tikzpicture}%
  }
  \caption{Localization Error at $z =$ \SI{3}{m}: Multipath vs. No Multipath}
  \label{fig:MP}
\end{figure}

\section{Conclusion}\label{sec:conclusion}
In this study, we explored and advanced the field of indoor localization algorithms, with a specific focus on deep learning methods, including the implementation of an innovative \emph{LSTM-PSO Hybrid} method. Our goal was not only to enhance the reliability of existing algorithms but also to assess their performance in realistic scenarios.

Both approaches demonstrated an extremely high level of reliability comparable to established methods, surpassing them by providing a much higher certainty that the estimation is close to the ground truth. LSTM has an accuracy in the centimeter range, while the hybrid approach showed an accuracy below the millimeter level maintaining extremely high reliability. 
The \emph{LSTM-PSO Hybrid} method emerges as a promising solution, showcasing a balanced fusion of the reliability inherent in neural networks and the high-resolution capabilities of optimization techniques. Compared with other techniques mentioned in \Cref{sec:sart} that also use AI, our study achieved higher accuracy. 
However, it is essential to recognize the need for further research regarding the challenge posed by multipath effects. Despite this current limitation, our study underscores the robustness of deep-learning methods, particularly the LSTM, revealing its efficacy across various practical scenarios, including the multipath effect,  which is lacking in current literature. This highlights the potential of this type of approach as a strong candidate for further exploration and development in the pursuit of techniques for RIS-based localization. 

Considering a practical system implementation, the methods show both strong aspects as well as some challenges. 
Our methods deliver reliable results in challenging real-life scenarios and excel in terms of computational efficiency, highlighted by the NN. This offers a significant advantage over traditional brute-force approaches when addressing large search areas. On the other hand, employing Deep Learning to address this challenge requires environment fingerprinting for model training, which may be a slow process, contingent upon the data collection method and environmental conditions. The model is robust to dynamic environments given adequate training data. However, like all machine learning models, it would need retraining if the environment undergoes major changes not seen during initial training.

\bibliographystyle{IEEEtran}
\bibliography{refs}

\end{document}